\documentclass [twocolumn]{article}
\usepackage{amsmath}
\usepackage{amsfonts}
\usepackage{amssymb}
\usepackage{graphicx}
\usepackage{float}
\usepackage{color}

\begin{document}
\title{Upper Bound on Singlet Fraction of Two Qubit Mixed Entangled States}
\author{Satyabrata Adhikari \thanks{satya@iitj.ac.in}, Atul Kumar \thanks{atulk@iitj.ac.in} \\\
Centre for System Science \\\ Indian Institute of Technology
Jodhpur, Jodhpur-342011, Rajasthan, India\\} \maketitle

\begin{abstract}
We demonstrate the possibility of achieving the maximum possible
singlet fraction using a entangled mixed two-qubit state as a
resource. For this, we establish a tight upper bound on singlet fraction and show that the maximal singlet fraction
obtained in \cite{Verstraete} does not attain the obtained upper
bound on the singlet fraction. Interestingly, we found that the required upper bound can in fact be achieved using local filtering operations.

\end{abstract}
\maketitle

\section{Introduction}
Quantum entanglement \cite{Einstein} has been used as an efficient
resource for several quantum communication protocols such as
teleportation, dense coding, and cryptography
\cite{Bennett1,Bennett2,Gisin}. The existence of long range
quantum correlations between entangled qubits allows the use of
such systems for information transfer in communication protocols.
In general, if a state is maximally entangled then the optimal
success of a communication protocol is a certainty. However, in
real experimental set-ups it is difficult to prepare a pure
maximally entangled state due to the difficulty in handling the
multi-particle quantum systems in terms of interaction with
environment, control, and preserving the necessary quantum
coherence \cite{Zanardi}. This leads to a situation where one may
have to deal with mixed entangled resources for quantum
information processing. This raises the question of usefulness of
such mixed entangled systems for efficient quantum information
processing. Verstraete and Verschelde \cite{Verstraete} have shown
that any entangled two-qubit mixed state can be used as a resource
for quantum teleportation using certain trace preserving local
operations and classical communications. The measure of usefulness
of any two-qubit mixed entangled resource $(\rho)$ for quantum teleportation is given by
teleportation fidelity, i.e. $f_{T}=\frac{2F+1}{3}$, where $F$ is the singlet fraction of the entangled resource defined as $F(\rho ) = \max _U \left\langle \psi  \right|U^\dag  \rho U\left| \psi  \right\rangle$ and $\left| \psi  \right\rangle  = \frac{1}{{\sqrt 2 }}\sum\limits_{i = 0}^1 {\left| {ii} \right\rangle}$ \cite{Horodecki}. For $F > \frac{1}{2}$, quantum teleportation is always successful and if the singlet fraction of the underlying state is unity then in principle perfect teleportation can be achieved. Hence for a state to be useful as a resource, the
singlet fraction or the teleportation fidelity must be maximized.
In this Letter, we address the following question: given an
entangled mixed state of two qubits as a resource for quantum
teleportation, what is the upper bound on singlet fraction that
can be achieved using all possible filtering operations? Our
results show an interesting observation that the maximal singlet
fraction obtained in \cite{Verstraete} can still be increased to
coincide with the required upper bound on singlet fraction. \par
We analyze the upper bound on the singlet fraction (and thus the
upper bound on teleportation fidelity) that can be achieved using
a two qubit entangled mixed state. We call this bound as Dembo's
bound on singlet fraction for any given two-qubit mixed state.
Although Verstraete {\em et al} have found that any two-qubit
mixed entangled state can be used to achieve $F > \frac{1}{2}$
using optimal local operations and classical communication, our
analysis in this Letter demonstrates that the singlet fraction or teleportation
fidelity obtained in \cite{Verstraete} does not achieve the
Dembo's upper bound on singlet fraction. We further demonstrate
that one can in fact obtain the optimal singlet fraction in
coherence with the Dembo's bound by applying additional filtering
operations. In general, our results show that for any state with
$F>1/2$, the Dembo's bound on singlet fraction can be successfully
achieved. The increase in singlet fraction using local operations
and hence the increase in teleportation fidelity will have a
significant impact on quantum information processing. In
principle, this will allow one to use even a mixed state to
achieve successful teleportation with optimal fidelity
\cite{Taketani}. Furthermore, our results may also release the
constraints on the experimental set-ups to prepare a pure
maximally entangled state for efficient and optimal quantum
teleportation. \par Verstraete {\em et al} have addressed the
problem of maximal achievable singlet fraction of entangled
two-qubit mixed states optimized over all trace preserving local
operations and classical communications. Our study is motivated by
the fact that the optimal singlet fraction $(F^{*})$ of the
filtered state obtained in \cite{Verstraete} does not exceed the
value of $2/3$. At this juncture, we would like to raise the
question of analyzing the upper bound on the singlet fraction of
the filtered state. Precisely, we are interested in analyzing
whether the value $F^{*}$ can be increased to coincide with the
upper bound on singlet fraction. In order to discuss the
importance of our results, we now proceed to establish a relation
between the upper bound on the singlet fraction and eigen values
of any Hermitian-positive semidefinite density operator. \par In
quantum information processing, eigenvalues play an important role
not only in shedding light on many physical aspects of quantum
systems but also for analyzing many essential features such as
entanglement, discord, communication, and security. For higher
dimensional systems, evaluating the exact maximum or minimum
eigenvalue of a Hermitian positive-definite matrix is not easy,
but its upper or lower bound suffices. To proceed further with our
analysis we state the following theorem \cite{Fang}: For any real
$n \otimes n$ matrix $C$ and a positive semidefinite operator $B$,
the following inequality holds
\begin{equation}
\lambda _1 (\bar C)tr(B) \le Tr(CB) \le \lambda _n (\bar C)tr(B)
\end{equation}
where $\bar C = \frac{{C + C^T }}{2}$, $C^{T}$ is the transpose of matrix $C$, $\lambda _n (\bar C)$ is the $nth$ eigen value of the matrix $\bar C$,
and $\lambda_{1} \leq \lambda_{2} \leq \lambda_{3} ..... \leq \lambda_{n}$. \\
For  $C = \frac{I_{4}}{2} - X^\Gamma$ and $B=\rho$, Eq. (1) can be re-expressed as
\begin{eqnarray}
\lambda _1 \left( {\frac{I_{4}}{2} - X^{\Gamma}} \right) \le F^{*} &=& tr\left[\left(\frac{I_{4}}{2}-X^{\Gamma}\right)\rho\right] \nonumber \\
&\le& \lambda _4 \left( {\frac{I_{4}}{2} - X^{\Gamma}} \right)
\end{eqnarray}
where $X = \left( {A \otimes I_{2}} \right)\left| \psi  \right\rangle ^ -  \left\langle \psi  \right|^ -  \left( {A^\dag   \otimes I_{2}} \right)$, $X^{\Gamma}$ is partial transpose of $X$,
$\left| \psi  \right\rangle ^ -   = \frac{1}{{\sqrt 2 }}\left\lceil {\left| {01} \right\rangle  - \left| {10} \right\rangle } \right\rceil$,
and $A$ represents the filter. Although the functional form of the upper bound of $F^{*}$ is optimal for the above inequality,
it depends on the state parameter and hence, must have a maximum achievable value for every particular value of the state parameter.
This value would be provided by Dembo's bound \cite{Dembo} stated below as: \\
For any $n \otimes n$ Hermitian positive semi-definite operator $R_{n}$ with eigen values $\lambda_{1} \leq \lambda_{2} \leq \lambda_{3}
..... \leq \lambda_{n}$,
Dembo's bound can be given by
\begin{eqnarray}
& &\frac{{c + \eta _1 }}{2} + \sqrt {\frac{{\left( {c - \eta _1 } \right)^2 }}{4} + b^* b}  \le \lambda _n (R_{n}) \nonumber \\ &\le& \frac{{c + \eta _{n - 1} }}{2} +
\sqrt {\frac{{\left( {c - \eta _{n - 1} } \right)^2 }}{2} + b^* b}
\end{eqnarray}
where $R_n  = \left( {\begin{array}{*{20}c} {R_{n - 1} } & b  \\
{(b^* )^T } & c  \\ \end{array}} \right)$, $\eta_{1}$ is the lower
bound on the minimal eigenvalue of $R_{n-1}$, $\eta_{n-1}$ is the
upper bound on the maximal eigen value of $R_{n-1}$, and $b$ is an
eigenvector of dimension $(n-1)$. Using Dembo's bound Eq. (2) can
be re-expressed as
\begin{eqnarray}
& &\lambda _1 \left( {\frac{I_{4}}{2} - X^{\Gamma}} \right) \le
F^{*}\le \lambda _4 \left( {\frac{I_{4}}{2} - X^{\Gamma}} \right) \nonumber \\
&\le& \frac{{c + \eta _3 }}{2} + \sqrt {\frac{{\left( {c - \eta _3 }
\right)^2 }}{2} + b^* b}
\end{eqnarray}
Therefore, the upper bound on optimal singlet fraction is
\begin{equation}
F^{*}_{D}=\frac{{c + \eta _3 }}{2} + \sqrt {\frac{{\left( {c - \eta _3 } \right)^2 }}{2} + b^* b}
\end{equation}
For the family of states given by
\begin{equation}
\rho(F)=F\left|\psi\right\rangle\left\langle \psi  \right|+(1-F) \left|01\right\rangle\left\langle  01\right| ;{\rm{    }} F \geq \frac{1}{3},
\end{equation}
the upper bound on singlet fraction $F^{*}_{D}$ is
\begin{equation}
\begin{array}{l}
 F^*_{D} [\rho (F)] = \frac{2-F}{4(1-F)} ;{\rm{    }}\frac{1}{3} \le F \le \frac{2}{3} \\
 F^*_{D} [\rho (F)] = F;{\rm{    }}F \ge \frac{2}{3}, \\
\end{array}{1}
\end{equation}
The value of $F^{*}_{D}$ obtained in Eq. (7) is the optimal value of singlet fraction that can be achieved for the family of states represented by Eq. (6).
Verstraete and Verschelde have shown that using trace preserving optimal local operations, the maximal achievable singlet fraction $F^{*}$ for the
family of states given in Eq. (6) is
\begin{equation}
\begin{array}{l}
 F^{*} [\rho (F)] = \frac{1}{2}\left[ {1 + \frac{{F^2 }}{{4(1 - F)}}} \right];{\rm{    }}\frac{1}{3} \le F \le \frac{2}{3} \\
 F^{*} [\rho (F)] = F;{\rm{    }}F \ge \frac{2}{3}, \\
\end{array}
\end{equation}
Eq. (7) and (8) describe an interesting result that $F^{*}$ does
not achieve the upper bound on singlet fraction given by
$F^{*}_{D}$. A comparison of maximal singlet fraction $F^*$
obtained after performing the filtering operations and $F^{*}_{D}$
obtained using Dembo's bound is given in Fig. (1) and clearly
demonstrates that the optimal singlet fraction $F^*$ is always
less than the optimal singlet fraction $F^*_{D}$ for $\frac{1}{3}
\le F \le \frac{2}{3}$. In general, we will always find that
$F^{*} \leq F^{*}_{D}$.
\begin{figure}
\includegraphics [width=0.45\textwidth] {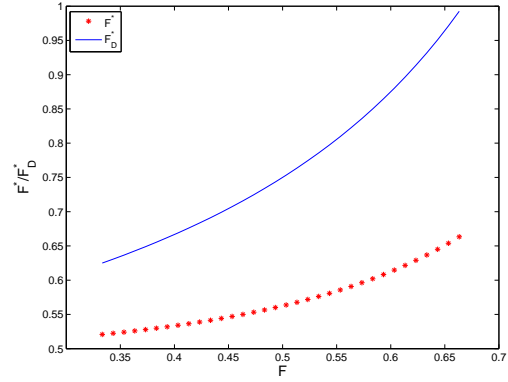}
\caption {Comparison of maximal singlet fraction
$F^{*}$ and upper bound on singlet fraction $F_{D}^{*}$ obtained
using Dembo's bound}
\end{figure}
Dembo's upper bound on singlet fraction is obtained using mathematical rigour. However, we still need to find a way to obtained this bound experimentally
i.e. would it be possible to increase the value of optimal singlet fraction performing local operations and classical communication on the filtered state
i.e. can we achieve the upper bound of singlet fraction given by Dembo's bound? Surprisingly, our results show that the bound is indeed achievable. \par
In order to enhance the value of optimal singlet fraction $F^*$, we perform another filtering operation on the filtered state such that the singlet
fraction of the output state can be given as
\begin{equation}
F^{*}_{opt}=p F^{*}(\rho_{f})+(1-p)F(\rho_{f})
\end{equation}
where $p$ is the success probability multiplied with the optimal singlet fraction of the state coming out of the second filter.
If we define $1-p=p_{AB}$ where $p_{AB}$ denotes the success probability of first filter (for details, please refer to \cite{Verstraete}), then for $F(\rho_f) = \left\langle \psi  \right|\rho _f \left| \psi  \right\rangle$
 where $\rho _f  = \frac{(A \otimes I)\rho  \left( {A \otimes I} \right)^\dag}{p_{AB}}$, $F^{*}_{opt}$ can be re-expressed as
\begin{equation}
F^{*}_{opt}=(1-p_{AB})F^{*}(\rho _f)+tr\left[(A \otimes I)\rho(A^\dag \otimes I)\left|\psi\right\rangle\left\langle \psi  \right|\right]
\end{equation}
The definition of $p_{AB}$ suggests that for success probability $p$ to be high, $p_{AB}$ must be minimized. Moreover, the minimum value of $p_{AB}$
should be chosen in such a way that the value of singlet fraction for the second filter must not exceed Dembo's bound. Hence, the minimum value of
$p_{AB}$ would be
\begin{equation}
p_{AB}^{min} = 1-\frac{F^{*}_{D}-tr\left[(A \otimes I)\rho(A^{T} \otimes I)\left|\psi\right\rangle\left\langle \psi  \right|\right]}{F^{*}(\rho_{f})}
\end{equation}
Using Eq. (10) and (11), we have
\begin{equation}
F^{*}_{opt}= F^{*}_{D}
\end{equation}
Eq. (12) clearly shows that one can achieve the maximum singlet fraction equals to Dembo's bound by using the filtering operations twice. \par
As an illustration, we use the family of states given by Eq. (6) to calculate the $F^{*}_{opt}$. In this case, Eq. (11) will become
\begin{equation}
p_{AB}^{min} = \frac{F^{2}}{2(1-F)(2-F)},
\end{equation}
and hence
\begin{equation}
F^{*}_{opt} = \frac{2-F}{4(1-F)},
\end{equation}
which is equal to $F^{*}_{D}$.\\
Hence, applying the filter twice will always result in achieving
the upper bound on singlet fraction for any two-qubit mixed
density operator. However, achieving this upper bound on singlet fraction will lead to the decrease in success probability of the first filter. Nevertheless, the second filter always compensates for the decrease in success probability of the first filter by attaining the Dembo's bound on singlet fraction.   \par
In summary, we have established a relation
between Dembo's upper bound and singlet fraction (and hence with
teleportation fidelity) of a mixed two-qubit entangled state. This
relation is used to demonstrate that any two-qubit mixed entangled
state can be used as a resource to achieve maximum possible
teleportation fidelity. It was found earlier that the the trace
preserving local operations always yield a teleportation fidelity
larger than $2/3$ if the original state is entangled. However, we
found that the maximal fidelity obtained earlier can be increased
with additional local operations with certain non-zero
probability. It would be interesting to find an example where the
optimal fidelity after the operation of a single filter can
coincide with Dembo's bound.

\end{document}